# Scale-dependent physical constraints on active intracellular fluctuations

**Yuika Ueda[1], Outa Nakashima[1], Takumi Saito[1], and Shinji Deguchi[1,2,3,*]**

[1] Division of Bioengineering, Graduate School of Engineering Science, The University of Osaka

[2] Global Center for Medical Engineering and Informatics, The University of Osaka

[3] R$^3$ Institute for Newly-Emerging Science Design, The University of Osaka

[*] Corresponding author

Address: 1-3 Machikane-yama, Toyonaka, Osaka 560-8531, Japan

E-mail: deguchi.shinji.es@osaka-u.ac.jp

Phone: +81 6 6850 6215

ORCID: 0000-0002-0556-4599

## Abstract

Living cells exhibit nonequilibrium dynamics that shape intracellular processes across length scales, from nanoscale molecular assembly to the organization of macroscopic organelles. While dynamics at micrometer scales are known to be constrained by the actin meshwork at low frequencies, the physical principles governing active fluctuations at the nanoscale remain elusive. Here, we present an analytical framework integrating fluorescence correlation spectroscopy with nonequilibrium modeling to delineate the physical scaling of intracellular mechanics. Applying this framework to fibroblasts, we demonstrate that, in contrast to larger components, nanoscale active fluctuations remain prominent at high frequencies and are predominantly driven by local nonmuscle myosin II activity, establishing a distinct functional hierarchy in intracellular mechanics: local active forces promote rapid spatial exploration for nanoscale molecules, whereas macroscopic actin constraints ensure the structural stability required for larger molecular complexes and organelles. To integrate these scale-dependent behaviors within a single physical framework, we formulated a model that captures the transition of active fluctuations across length scales, revealing that the physical properties of the cytoplasm are governed by the balance between active driving forces and passive structural constraints. Furthermore, applying this model to cellular senescence reveals a reduction in nonequilibrium complexity associated with cytoskeletal rigidification. Thus, our findings bridge the dimensional gap between local molecular kinetics and macroscopic constraints, providing a fundamental physical basis for understanding the hierarchical organization of intracellular dynamics.

# 1. Introduction

The intracellular environment is a dynamic and mechanically active space where diverse biological processes are continuously coordinated (1). Rather than existing in thermal equilibrium, the cellular interior is driven by active forces generated through the hydrolysis of ATP by motor proteins and the remodeling of the cytoskeleton (2,3). These processes generate nonequilibrium fluctuations throughout the cytoplasm (4–7). A fundamental paradox in biophysics lies in the ability of cells to execute coordinated functions, such as intracellular transport and structural organization, while being constructed from components subjected to continuous stochastic fluctuations. Uncovering the physical principles that govern how cells harness these active fluctuations to maintain functional robustness across scales remains a critical challenge.

Efforts to address this dimensional challenge have advanced our understanding of active intracellular mechanics. Studies utilizing active microrheology and single-particle tracking with relatively large probes (~100 nm–1 μm) have demonstrated that probe dynamics are predominantly governed by the macroscopic physical constraints and nonlinear deformations of the actin meshwork (4–10). However, the physical principles governing active fluctuations at the nanoscale remain elusive. This lack of understanding represents a critical knowledge gap because individual proteins and molecular complexes operate at this scale. The primary obstacle to resolving this gap has been methodological. Traditional tracking techniques are limited by spatial and temporal resolution in capturing the rapid and minute displacements of nanoscale particles. Consequently, how nonequilibrium mechanical properties transition from larger-probe dynamics to the nanoscale intracellular environment remains a major unresolved question.

To bridge this dimensional divide, we develop a quantitative framework that integrates nonequilibrium physics with fluorescence correlation spectroscopy (FCS) to delineate the physical scaling of intracellular dynamics. Applying this framework to fibroblasts, we reveal that, in contrast to larger components, nanoscale active fluctuations remain prominent at high frequencies and are predominantly driven by local nonmuscle myosin II activity, suggesting a distinct functional hierarchy in intracellular mechanics: local myosin-driven forces enable rapid spatial exploration for nanoscale molecules, whereas larger-scale structural constraints ensure the stability of larger molecular complexes and organelles. To account for these scale-dependent behaviors, we formulated a theoretical model that captures the transition between localized motor-driven forces and macroscopic structural constraints.

We further applied our approach to cellular senescence, a physiological transition accompanied by profound mechanical and structural alterations. Senescent cells exhibit reorganization and stabilization of actin stress fibers that reinforce substrate anchorage (11,12), along with increased cytoplasmic stiffness and crowding that amplify active intracellular forces while reducing overall intracellular mobility (13). Our analysis reveals that senescence is characterized by a marked reduction in fluctuation diversity. While the overall magnitude of active forces increases during senescence, the underlying fluctuation modes become significantly simplified. These findings suggest that transitions in cellular state are accompanied by distinct changes in nonequilibrium dynamics. Together, our results support a hierarchical organization of intracellular mechanics underlying robust biological functions across intracellular scales.

## 2. Material and methods

### 2.1 Cell culture

Primary human foreskin fibroblasts (HFF1, ATCC) were cultured in high glucose DMEM (Wako) supplemented with 15% fetal bovine serum (Sigma Aldrich) and 1% penicillin–streptomycin solution (Wako). Cells were kept at 37 °C in a humidified incubator with 5% $CO_2$. The culture medium was changed once every three days, and cells were passaged at a split ratio of 1 to 3 when reaching 80% confluence. For cellular senescence experiments, HFF1 cells at passage 26 (p26), previously validated as senescent using multiple senescence markers (11,12), were compared with younger cells at passage 9 (p9).

### 2.2 Pharmacological treatments

For pharmacological treatments, cells were incubated with the indicated agents for 10 min before and throughout the measurements. Nonmuscle myosin II activity was inhibited using (-)-blebbistatin (Wako) at 10 or 20 μM, whereas calyculin A (Wako) was used at 5 nM to enhance nonmuscle myosin II activity. To examine the contribution of other motor proteins, cells were treated with the dynein inhibitor ciliobrevin D (GlpBio) or the kinesin inhibitor monastrol (LKT Laboratories), each at 10 or 20 μM. The role of the actin cytoskeleton was assessed using cytochalasin D (Wako), an actin polymerization inhibitor, at 10 μM and 20 μM, and jasplakinolide (Sigma-Aldrich), an actin filament stabilizer, at 100 nM. Microtubule polymerization was inhibited using nocodazole (Wako) at 10 or 20 μM. Combined actomyosin inhibition was performed using a mixture of (-)-blebbistatin and cytochalasin D at the indicated concentrations.

## 2.3 Plasmids and transfection

The fluorescent protein mClover2 was used as a tracer for intracellular fluctuation measurements. The mClover2-C1 plasmid (Addgene plasmid #54577) was used as the monomeric mClover2 construct. To examine the effect of tracer size, previously generated and validated tandem mClover2 constructs (14), in which two or three mClover2 units were linked in series, were used as 2-mClover2 and 3-mClover2, respectively. For myosin heavy chain overexpression, the mCherry-MYH9 plasmid was used (15). Cells were transfected with the plasmids 24 h after seeding using Lipofectamine LTX and Plus Reagent (Thermo Fisher Scientific) according to the manufacturer's instructions.

## 2.4 Fluorescence correlation spectroscopy

To probe intracellular dynamics, we utilized FCS, which analyzes fluctuations in fluorescence intensity from diffusing molecules within a microscopic observation volume. The statistical properties of these fluctuations were quantified using the fluorescence intensity time-autocorrelation function. FCS measurements were performed on a confocal laser scanning microscope (Olympus FV3000) equipped with a 60× oil-immersion objective. Live cells cultured on glass-bottom dishes for 24 h after transfection were maintained at 37 °C and 5% $CO_2$ using a humidified stage-top incubator. The pinhole was set to 200 μm, resulting in a three-dimensional Gaussian observation volume with a radial dimension of 0.211 μm and a structural parameter of 2.67. Detailed mathematical definitions of the autocorrelation function and fitting procedures used to extract the diffusion coefficient are provided in the Supplementary Information.

## 2.5 Quantification of nonthermal forces using a nonequilibrium model

To quantify active forces within cells, we developed a nonequilibrium physical framework based on the violation of the Einstein relation. In living cells, intracellular particles are subjected to both random thermal forces arising from molecular collisions and active nonthermal forces generated by molecular motors and cytoskeletal remodeling. We related the average particle velocity to the total driving force and used FCS-derived parameters, including the effective observation length, diffusion coefficient, and correlation time, to evaluate the degree of nonequilibrium in microscopic particle motion. The nonthermal force was calculated as follows:

$$f_A = \frac{l}{D\tau_B}\left(\frac{D}{\mu} - k_B T\right) \tag{1}$$

Here, $f_A$ represents the nonthermal force, which quantifies the degree of nonequilibrium. The parameter $l$ denotes the effective radial length of the FCS observation volume, corresponding to the radial dimension described above. The diffusion coefficient $D$ and correlation time $\tau_B$ were obtained by fitting the acquired fluorescence autocorrelation function to a three-dimensional diffusion model; detailed mathematical derivations are provided in the Supplementary Information. The coefficient $\mu$ denotes particle mobility, which was calculated using Stokes law with an assumed effective molecular radius of 1.2 nm based on the reported dimensions of GFP (16), and an intracellular macroscopic viscosity of 10 Pa·s. This viscosity represents a typical order of magnitude for the crowded cytoplasmic environment, where the actin meshwork restricts particle motion (5,17). The term $k_B T$, where $k_B$ is the Boltzmann constant ($1.38\times10^{-23}$ J/K) and $T$ is the absolute temperature (310 K), represents the thermal fluctuation component. By measuring the deviation of observed fluctuations from those predicted by purely thermal Brownian motion, the magnitude of effective nonthermal forces can be estimated. This approach allows evaluation of the intracellular nonequilibrium state from FCS-derived diffusion parameters without requiring external mechanical stimuli. Because the absolute value depends on assumptions regarding particle mobility and cytoplasmic viscosity, we primarily used this quantity to compare relative changes in intracellular nonequilibrium activity across experimental conditions.

## 2.6 Theoretical model for displacement power spectral density

To interpret the experimental data, we constructed a theoretical model incorporating cytoplasmic viscoelasticity and actomyosin dynamics. The displacement power spectral density $C(\omega)$ was defined as:

$$C(\omega) = \frac{S_{th} + S_{act}(\omega)}{k^2 + \gamma^2\omega^2} \tag{2}$$

$$S_{act}(\omega) = \sum_{i\in\{\text{fast,slow}\}} \frac{2\tau_i f_i^2}{1 + (\omega\tau_i)^2} \tag{3}$$

where $S_{th}$ denotes thermal noise, $k$ is the elastic cage stiffness, and $\gamma$ is the friction coefficient. Detailed derivations are provided in the Supplementary Information. The active noise spectrum $S_{act}(\omega)$ incorporates two characteristic time scales: a fast time scale ($\tau_{fast} \approx 0.01$ s), corresponding to individual force generation steps of nonmuscle myosin II motors (18), and a slow time scale ($\tau_{slow} \approx 1$ s), reflecting dynamic structural remodeling of the cross-linked actomyosin network (19,20).

## 2.7 Statistical analysis

All data were obtained from three independent experiments. The number of analyzed regions for each condition is indicated in the corresponding figure legends. Box plots show the median, mean, interquartile range, whiskers extending to 1.5 × IQR, and individual data points. Statistical significance was determined using a two-sided Welch's *t*-test. Asterisks indicate statistical significance: *$p < 0.05$, **$p < 0.01$, and ***$p < 0.001$.

# 3. Results

## 3.1 Myosin II activity drives intracellular nonthermal fluctuations

The obtained nonthermal force exceeded the force generated by individual molecular motors, suggesting that the measured fluctuations reflect collective nonequilibrium activity rather than isolated motor events. We therefore used this quantity as an effective index of intracellular active fluctuations and examined how it depends on cytoskeletal motor activity. To identify the molecular origin of the nonthermal force, active fluctuations were quantified using the framework defined by Eq. (1) (Fig. 1a). Specific inhibition of nonmuscle myosin II with (-)-blebbistatin at 10 or 20 μM significantly decreased the nonthermal force, whereas myosin heavy chain overexpression had the opposite effect (Fig. 1b). Treatment with calyculin A, a phosphatase inhibitor known to induce sustained nonmuscle myosin II hyperphosphorylation and rigid actomyosin cross-linking (21,22), also resulted in a significant decrease in the nonthermal force, suggesting that excessive actomyosin stiffening restricts the flexible force generation required to drive active fluctuations. Collectively, these results indicate that dynamic nonmuscle myosin II activity is a central driver of intracellular nonthermal fluctuations.

## 3.2 Kinesin and dynein have little effect on nonthermal fluctuations

To determine whether microtubule-based molecular motors contribute to the observed active forces, we examined the effects of inhibiting kinesin and dynein, which mediate intracellular transport along microtubules. Specific inhibition of dynein with ciliobrevin D or kinesin with monastrol at 10 or 20 μM resulted in no notable changes in the nonthermal force (Fig. 1c). These results suggest that the measured nonthermal fluctuations are independent of microtubule-based motor activity, distinguishing them from directed transport processes mediated by kinesin and dynein.

## 3.3 Actin network dynamics contribute to nonthermal fluctuations

To further identify cytoskeletal components involved in the nonthermal fluctuations, we examined the effects of perturbing actin and microtubule networks. Disruption of actin polymerization with cytochalasin D led to a significant reduction in the nonthermal force at both 10 and 20 μM (Fig. 1d). Stabilizing the actin filaments with jasplakinolide at 100 nM also resulted in a substantial decrease in nonthermal fluctuations. These findings suggest that both the structural integrity and dynamic turnover of the actin network are critical for sustaining the observed nonthermal fluctuations. In contrast, depolymerization of microtubules with nocodazole at 10 or 20 μM had no notable effect on the nonthermal force, consistent with the limited contribution of microtubule-based motor activity. To further evaluate whether actin and myosin II operate as an integrated system, cells were subjected to simultaneous inhibition of myosin II activity and actin polymerization. This concurrent disruption substantially decreased the nonthermal force (Fig. 1e), suggesting that the actomyosin network serves as a structural basis for nanoscale nonequilibrium fluctuations.

## 3.4 Size dependence of nonthermal fluctuations and the effect of spatial confinement

To investigate how the scale of intracellular probes affects the measured dynamics, the size dependence of the nonthermal force was evaluated using monomeric, dimeric, and trimeric mClover constructs, denoted as 1-mClover, 2-mClover, and 3-mClover, respectively. The magnitude of the nonthermal force substantially decreased as probe size increased from the monomer to the trimer (Fig. 2). Specific inhibition of nonmuscle myosin II using (-)-blebbistatin yielded no notable change in the nonthermal force for either 2-mClover or 3-mClover.

Interestingly, disruption of actin polymerization with cytochalasin D increased the nonthermal force for 2-mClover, in contrast to the reduction observed for 1-mClover. This result suggests that, for larger probes, release from spatial confinement imposed by the actin meshwork exerts a greater effect than the loss of the actomyosin-driven active force. The typical pore size of the intracellular actin meshwork has been estimated to be approximately 20–40 nm (7) or 30–100 nm (23). Considering that the hydrodynamic radius of a monomeric fluorescent protein is on the order of several nanometers (16), the multimeric constructs attain physical dimensions that become increasingly susceptible to steric restriction within this crowded meshwork. As the probe size increases in the dimeric construct, its motion becomes constrained by this physical barrier. Consequently, depolymerization of the actin meshwork relieves this confinement, leading to an overall increase in nonthermal fluctuations.

For the 3-mClover construct, cytochalasin D treatment resulted in a decrease in the nonthermal force, although the control group exhibited substantial variance. This broad distribution in the control data may reflect structural heterogeneity of the trimeric construct. Depending on whether the three connected domains adopt an extended linear conformation or a compact state, the effective probe may vary. Such structural variability is expected to alter the degree of transient confinement within the actin meshwork, thereby contributing to the wide distribution of the observed nonthermal forces.

## 3.5 Theoretical unification of scale-dependent physical behavior

How does the physical scale of observation dictate active dynamics within the same cytoplasm? This question arises from an apparent discrepancy between our nanoscale measurements and previous macroscopic studies. At the micrometer scale, corresponding to large structures and organelles, active fluctuations are predominantly restricted to low frequencies (typically below 10 Hz) (5). Furthermore, disruption of the actin meshwork increases fluctuations by relieving spatial confinement, while myosin II inhibition has little effect (7). In contrast, our nanoscale measurements reveal significant active fluctuations extending into high frequencies (typically exceeding 500 Hz). In this regime, disrupting either the actin meshwork or myosin II activity decreases active fluctuations. This scale-dependent difference can be explained by the elastic caging effect of the actin meshwork. When the probe size is substantially smaller than the characteristic mesh size, particles can escape this spatial confinement, allowing local active forces to dictate the dynamics. However, as the probe size approaches or exceeds the mesh dimensions, particle motion becomes increasingly constrained, rendering the dynamics dominated by the elastic cage rather than by active driving forces. To reconcile these scale-dependent behaviors, we constructed a theoretical model incorporating cytoplasmic viscoelasticity and actomyosin dynamics (Fig. 3a, Eqs. (2) and (3)).

Simulations varying the probe radius $a$ relative to the mesh size $x_i \approx 100$ nm (24,25) reproduced these contrasting behaviors. For a small probe ($a = 5$ nm), the elastic caging $k$ is negligible. The dynamics are primarily driven by local active forces, resulting in a substantial decrease in fluctuations upon either myosin II or actin inhibition across a broad high-frequency range (Figs. 3b, c), consistent with our experimental observations at the single-protein scale. In contrast, for a large probe ($a = 10$ μm), the elastic cage $k$ dominates the dynamics. Inhibiting myosin II has little effect, whereas actin disruption removes the spatial constraint imposed by the elastic cage. This release of confinement outweighs the loss of active force, thereby increasing the low-frequency fluctuations (Figs. 3d, e), consistent with previous macroscopic observations.

Mapping this dynamic balance yields a phase diagram of intracellular fluctuations (Fig. 3f). This theoretical framework also provides a physical explanation for the size-dependent reversal and broad variance observed in the multimeric mClover constructs (Fig. 2). As the probe size increases from 1-mClover toward 3-mClover, the construct approaches the mesh dimensions and begins to experience elastic confinement. The 3-mClover construct can adopt various conformations, ranging from relatively extended to compact states. This conformational diversity generates a broad distribution of effective probe sizes, leading to variable degrees of transient confinement within the actin meshwork. In the theoretical model, this variability corresponds to variability in the elastic constraint $k$ among individual probes, providing a physical rationale for the substantial variance observed in the 3-mClover experiments. Together, this scale transition shifts the dominant balance from active force-driven fluctuations to structurally restricted dynamics, seamlessly unifying our nanoscale findings with macroscopic intracellular mechanics.

## 3.6 Simplification of fluctuation modes during cellular senescence

Cellular senescence is accompanied by profound alterations in cytoskeletal architecture, typically characterized by the accumulation of rigid actin stress fibers (11). To investigate how this structural rigidification affects nanoscale active dynamics, intracellular fluctuations were compared between young cells (passage 9, p9) and senescent cells (passage 26, p26) (Fig. 4a). Quantification of nonthermal forces revealed a significant increase in this effective index in senescent cells (Fig. 4b). This increase is consistent with the enhanced actomyosin contractility required to maintain the tense, spread structure of senescent cells. Despite this increase in total force, the fluctuation dynamics became less diverse. Active fluctuation modes were evaluated by extracting correlation time scales from the autocorrelation profiles. In young cells, a substantial fraction of the recorded profiles (42.2%) required double-$\tau$ fitting, indicating the coexistence of multiple active fluctuation modes. In contrast, senescent cells showed a pronounced shift toward single-$\tau$ fitting, with 83.1% of profiles classified as single mode and the proportion of the double-mode decreasing to 13.5% (Fig. 4c). These observations suggest that senescence-associated structural rigidification is linked to the restriction of active intracellular fluctuation modes.

## 4. Discussion

The intracellular environment is not merely a viscous fluid in thermal equilibrium, but an active, mechanically structured space driven by continuous energy dissipation. In this study, we established a quantitative framework combining FCS with nonequilibrium theoretical modeling to resolve a central question in intracellular mechanics, the scale-dependent transition of active fluctuations. We found that nanoscale active dynamics are fundamentally distinct from macroscopic behaviors; whereas macroscopic fluctuations are restricted by the structural properties of the actin meshwork (7,26), nanoscale dynamics are characterized by high-frequency, localized active forces generated by actomyosin activity (4–6,27). This physical divergence suggests a functional hierarchy within the cytoplasm. Nanoscale fluctuations, corresponding to the dimensions of individual proteins, likely facilitate rapid spatial exploration and local structural remodeling. In contrast, macroscopic dynamics at the micrometer scale, involving organelles and the cytoskeletal matrix, resist high-frequency perturbations, providing the structural stability necessary for large scale mechanical propagation and coordinated architectural changes. We demonstrated that this scale dependent physical rationale provides a mechanistic framework for understanding complex biological transitions. By applying this model to cellular senescence, we found that cellular aging is associated with a profound simplification of active fluctuation modes, accompanied by structural rigidification (11,28,29). Together, our findings bridge the dimensional gap between local molecular kinetics and macroscopic constraints, revealing the hierarchical organization of intracellular mechanics that shapes cellular states.

While previous continuum models and polymer network theories have successfully described the macroscopic elasticity of the cytoskeletal meshwork (30,31), these frameworks typically approximate the intracellular environment as an effective medium. Such assumptions often overlook size-dependent anomalous transport that occurs when probe dimensions approach the mesh size (32). Our theoretical framework addresses this issue by explicitly separating active driving forces from passive structural constraints, accounting for the physical transition observed as probe size increases. Previous studies at micrometer scale demonstrated that active fluctuations are tightly confined by the elastic caging effect of the actin meshwork, where structural disruption increases fluctuation amplitude by releasing steric hindrance (7). Conversely, our nanoscale data establish that for the single-protein scale (1-mClover), the elastic caging $k$ is negligible, and dynamics are dictated by local active forces $S_{act}$.

Our multimeric mClover constructs capture the progression of this physical crossover. As the probe size increases to the 2-mClover construct, the mechanical behavior shifts toward the macroscopic regime; myosin II inhibition yields no significant change, whereas actin disruption leads to an increase in fluctuations. These observations suggest that the 2-mClover probe has already reached a critical dimension where the release of elastic caging outweighs the loss of active driving forces. Furthermore, this structural transition is amplified in the 3-mClover construct, which exhibits substantial data variance alongside the macroscopic response. This variance may reflect the heterogeneous nature of the intracellular environment. Unlike spherical beads, the 3-mClover construct possesses structural heterogeneity, existing in various conformations ranging from extended linear chains to compact states. These diverse shapes result in variable degrees of transient entanglement with the actin filaments, leading to variability in the elastic constraint $k$. Together, these findings suggest that multimeric probes can serve as sensitive sensors of the local meshwork architecture, seamlessly bridging the gap between discrete molecular kinetics and structurally restricted macroscopic mechanics.

The simplification of active fluctuation modes in senescent cells represents a mechanical signature of aging, characterized by a transition from heterogeneous exploration to structural restriction. Our findings demonstrate that while this effective index increases during senescence, consistent with previous studies reporting heightened actomyosin contractility in senescent cells (13), the underlying dynamics become predominantly monotonic. This phenomenon can be physically interpreted through the increase in the elastic constraint $k$ within our theoretical framework. In young cells (passage 9), the flexible actin meshwork allows nanoscale probes to experience a broad spectrum of active forces arising from both discrete myosin steps and local network remodeling, resulting in heterogeneous fluctuation modes. In contrast, the accumulation of actin stress fibers in senescent cells (passage 26) rigidifies the intracellular environment (31). This structural transformation effectively shifts the senescent cytoplasm toward a structurally restricted regime, where increased elastic caging traps local macromolecules and suppresses the diversity of active modes. This physical simplification extends beyond macroscopic stiffening and may impose constraints on the dynamic plasticity required for flexible cellular adaptation. The enhanced elastic caging restricts the network remodeling required for cells to reorganize their architecture in response to external mechanical or chemical cues. Therefore, we propose that a hallmark of cellular aging manifests physically not merely as the accumulation of tension, but as the reduction of nonequilibrium complexity and a corresponding loss of the dynamic degrees of freedom essential for physiological adaptability.

In conclusion, our study quantifies nanoscale active fluctuations through an analytical framework that combines FCS with nonequilibrium theoretical modeling. This integrated approach bridges the dimensional gap between local molecular kinetics and macroscopic intracellular mechanics. By resolving the scale-dependent paradox of active fluctuations, we demonstrate that the physical properties of the cytoplasm are governed by the balance between active driving forces and passive structural constraints at specific observational scales. This hierarchical organization of fluctuations supports cellular robustness by balancing local agility with global stability. The application of this model to cellular senescence underscores its physiological relevance, showing that the loss of physical diversity can serve as a quantitative metric of cellular states. While our current theoretical approach focuses on the physical constraints of the meshwork, future studies exploring the interplay between biochemical signaling pathways and mechanical fluctuations will further clarify how these active networks are dynamically regulated. This framework may also help decode the mechanical signatures of other biological transformations involving cytoskeletal reorganization, such as malignant transformation and stem cell differentiation. Ultimately, delineating how living cells spatiotemporally harness and restrict nonequilibrium fluctuations provides a physical basis for understanding the dynamic adaptability and functional robustness of biological systems.

## Acknowledgments

We thank Pirawan Chantachotikul for valuable guidance and support in cellular senescence research. YU is supported by the Japan Society for the Promotion of Science (JSPS). This study was supported in part by JSPS KAKENHI grants (23H04929, 23H04928, and 26KJ0217) and JST ACT-X (JPMJAX25L2).

## Declaration of interests

The authors declare no competing interests.

# Figures

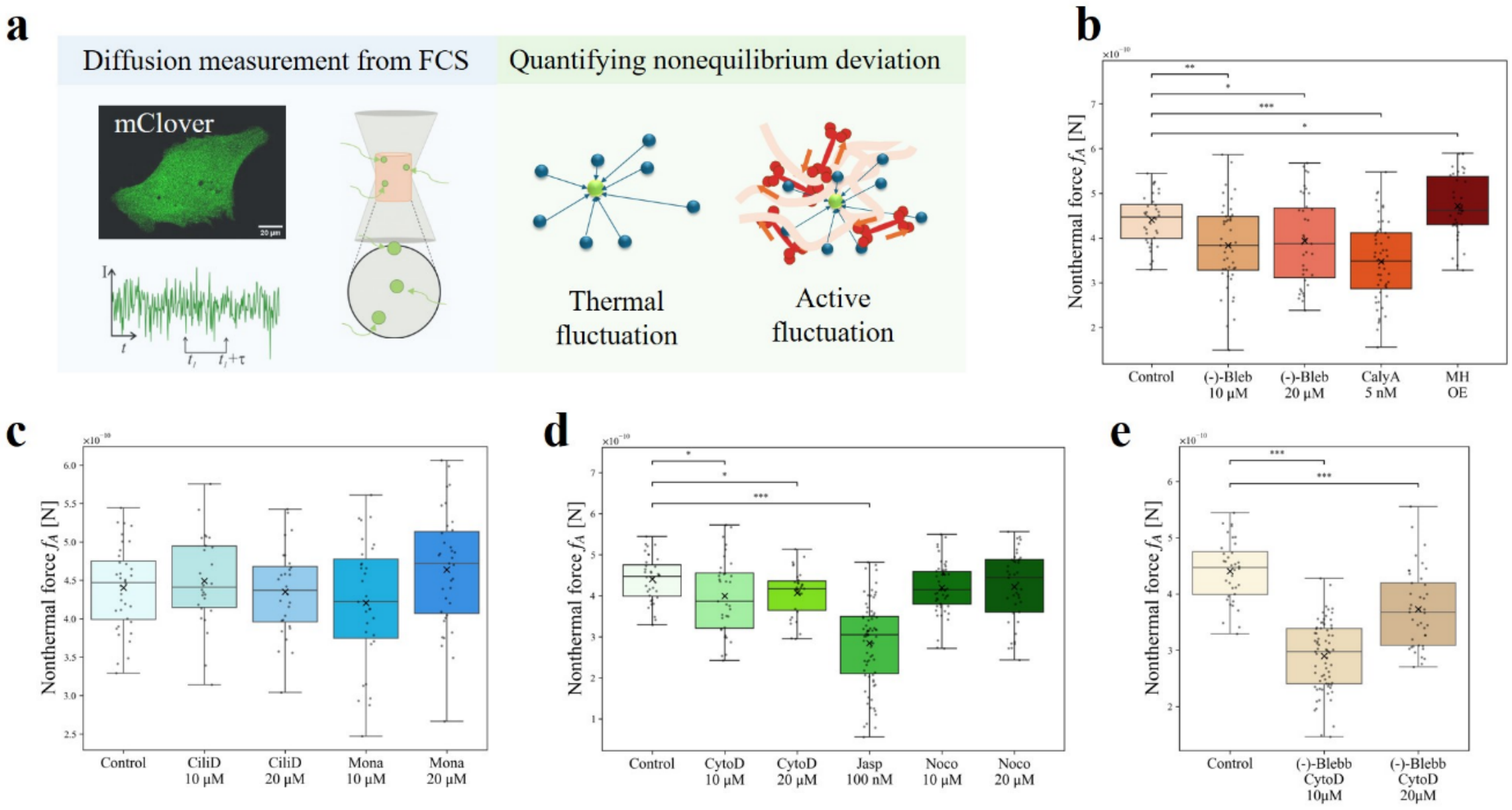

**Fig. 1 Molecular origins of active intracellular fluctuations.**

(a) Schematic of the framework used to quantify intracellular active fluctuations by FCS. Nonthermal force, an indicator of nonequilibrium states, was calculated from diffusion parameters using Eq. (1). (b) Effects of modulating myosin II activity on nonthermal forces. Cells were treated with (-)-blebbistatin or calyculin A, or subjected to myosin heavy chain overexpression ($n$ = 36, 44, 39, 52, and 38 regions for control, 10 μM (-)-blebbistatin, 20 μM (-)-blebbistatin, calyculin A, and myosin heavy chain overexpression, respectively). (c) Effects of ciliobrevin D and monastrol on nonthermal forces ($n$ = 36, 26, 28, 27, and 34 regions for control, 10 μM ciliobrevin D, 20 μM ciliobrevin D, 10 μM monastrol, and 20 μM monastrol, respectively). (d) Effects of cytochalasin D, jasplakinolide, and nocodazole on nonthermal forces ($n$ = 36, 26, 27, 75, 49, and 41 regions for control, 10 μM cytochalasin D, 20 μM cytochalasin D, jasplakinolide, 10 μM nocodazole, and 20 μM nocodazole, respectively). (e) Effect of combined (-)-blebbistatin and cytochalasin D treatment on nonthermal forces ($n$ = 36, 71, and 41 regions for control, 10 μM combined inhibition, and 20 μM combined inhibition, respectively).

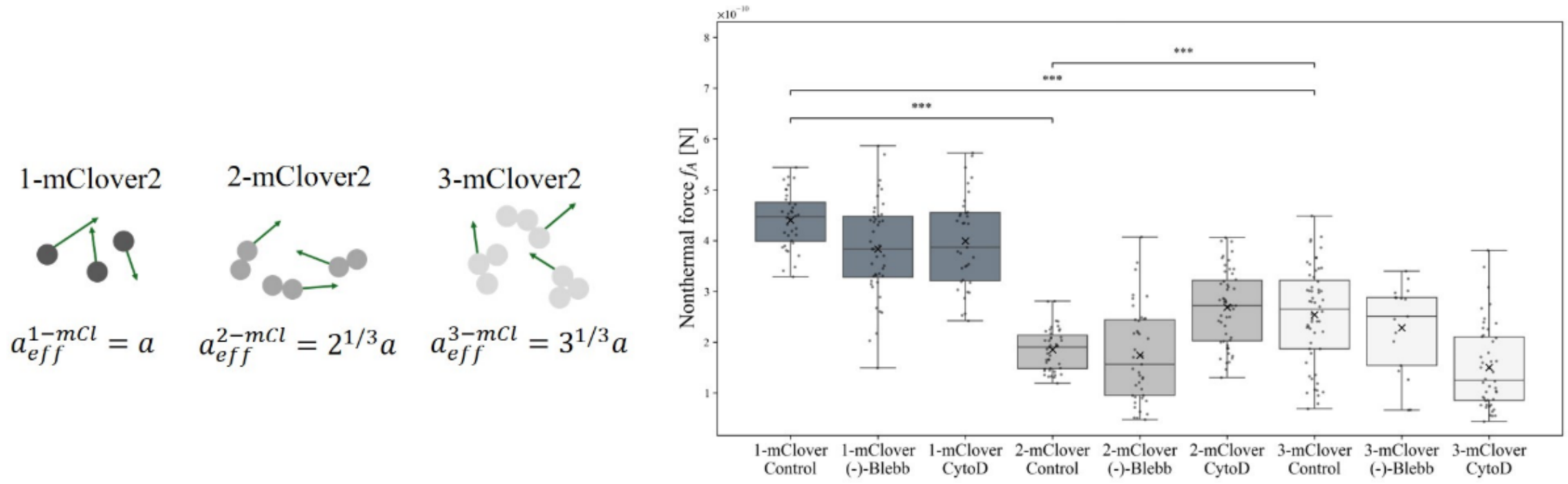

**Fig. 2 Size dependence and spatial confinement of nonthermal forces.**

Quantification of nonthermal forces across multimeric probes to evaluate probe size dependence and spatial confinement: 1-mClover ($n$ = 36, 44, and 26 regions for control, (-)-blebbistatin, and cytochalasin D treatments, respectively), 2-mClover ($n$ = 45, 41, and 53 regions for control, (-)-blebbistatin, and cytochalasin D treatments, respectively), and 3-mClover ($n$ = 56, 17, and 47 regions for control, (-)-blebbistatin, and cytochalasin D treatments, respectively).

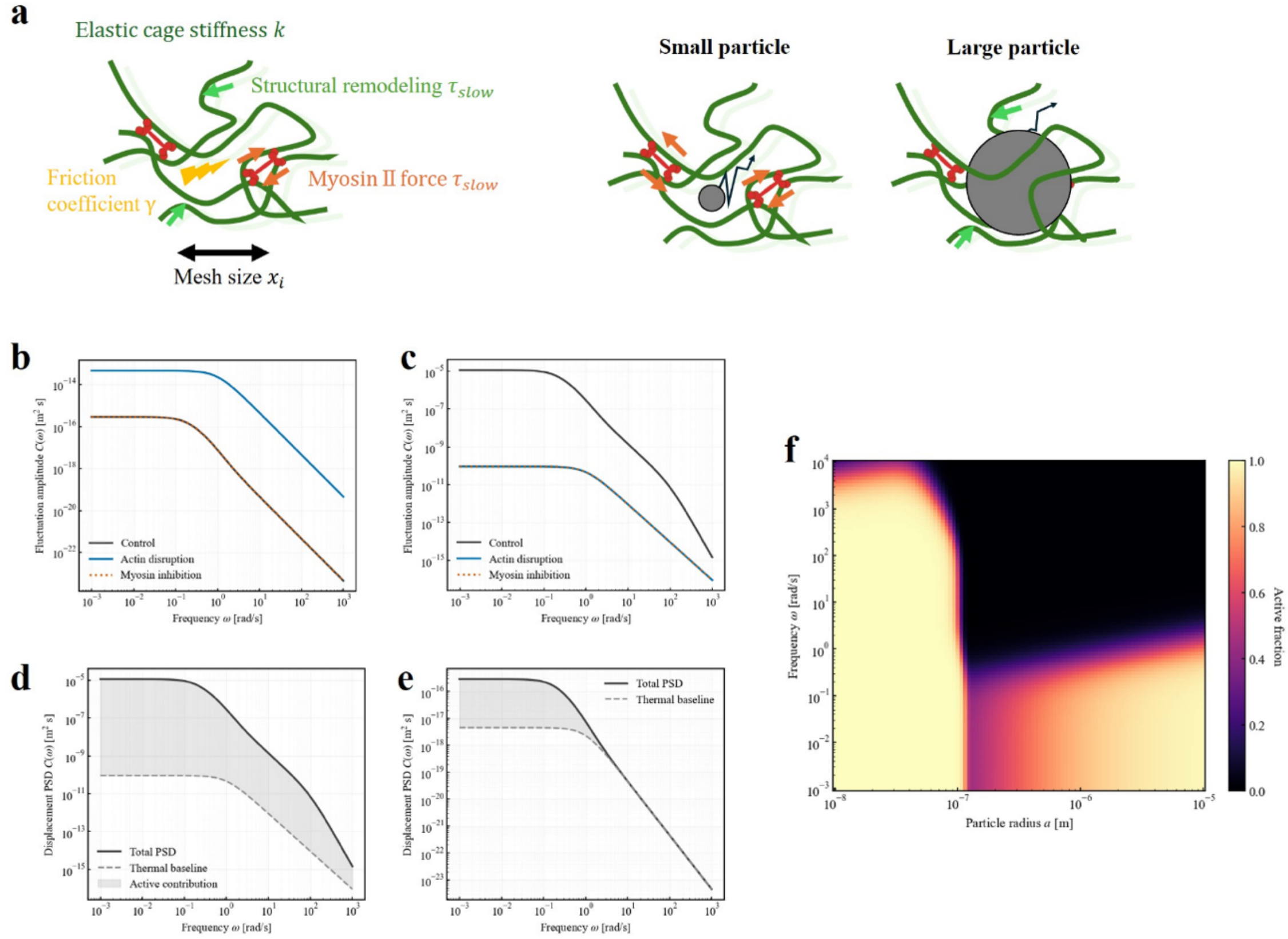

**Fig. 3 Elucidation of scale-dependent mechanics through a theoretical model.**

(a) Schematic of the theoretical model incorporating cytoplasmic viscoelasticity and actomyosin dynamics. The displacement power spectral density $C(\omega)$ of intracellular fluctuations is defined as the sum of thermal noise $S_{th}$ and an active noise spectrum $S_{act}(\omega)$ (Eqs. (2) and (3)). This active spectrum accounts for two time scales: individual force generation steps by myosin II motors ($\tau_{fast}$) and dynamic structural remodeling of the cross-linked actomyosin network ($\tau_{slow}$). Here, $k$ represents the elastic cage stiffness, and $\gamma$ is the friction coefficient. (b, d) Simulated effects of varying probe radius $a$ relative to mesh size $x_i$ on the power spectral density of nonthermal forces. Power spectral densities are shown for a small particle ($a$ = 5 nm, (b)) and a large particle ($a$ = 10 μm, (d)). (c, e) Simulated impacts of myosin inhibition and actin meshwork disruption on the power spectral density. For small particles (c), the elastic caging $k$ is negligible, and therefore both perturbations reduce active fluctuations across a broad frequency range. Conversely, for large particles (e), the elastic cage k is dominant, and the loss of steric hindrance due to actin disruption elevates low-frequency fluctuations. (f) Phase diagram of intracellular fluctuations showing regimes dominated by active force or elastic caging in relation to probe size and active-force frequency.

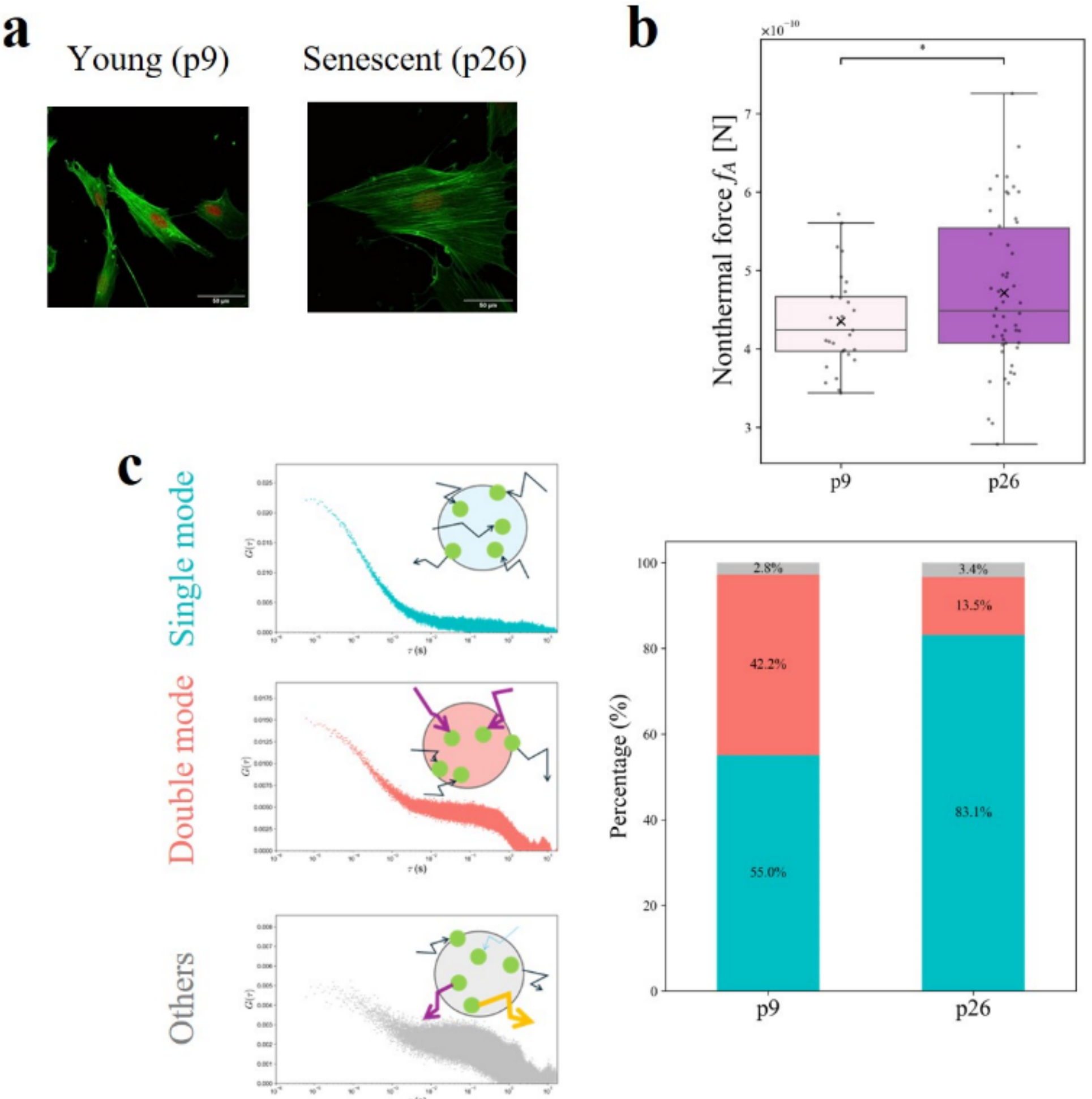

**Fig. 4 Cellular senescence simplifies intracellular fluctuation modes.**

(a) Schematic comparing intracellular fluctuations between young cells (passage 9, p9) and senescent cells (passage 26, p26). (b) Quantification of nonthermal forces in young and senescent cells ($n$ = 30 and 50 regions for p9 and p26, respectively). The effective index of nonthermal fluctuations increased significantly in senescent cells, consistent with heightened actomyosin contractility required to maintain the tense, spread morphology characteristic of senescent cells. (c) Analysis of active fluctuation modes in young and senescent cells. The proportions of single modes (single) and double modes (double), indicating the coexistence of multiple active fluctuation modes, are shown based on correlation time scales extracted from FCS autocorrelation profiles. In young cells, 42.2% of the recorded profiles required double modes. In contrast, active dynamics are simplified in senescent cells, converging to a single mode in 83.1% of cases, with the proportion of double modes decreasing to 13.5%.

# Supporting Information

# of Ueda et al., "Scale-dependent physical constraints on active intracellular fluctuations"

## Nonequilibrium model for analyzing nonthermal forces

To quantify nonthermal forces within cells, we developed a nonequilibrium physical framework based on the violation of the Einstein relation. The time evolution of the intracellular particle density $\rho$ at position $x$ is described by

$$\frac{\partial \rho}{\partial t} = u\frac{\partial \rho}{\partial x} + D\frac{\partial^2 \rho}{\partial x^2} \qquad (s1-1)$$

where $u$ is the average particle velocity, $t$ is time, and $D$ is the diffusion coefficient, assuming that the intracellular region under consideration is sufficiently large compared with the size of individual particles. At steady state with negligible macroscopic particle flow, $\rho$ remains constant over time, allowing the steady-state density $\rho_{st}(x)$ to be described as

$$\rho_{st}(x) = \rho_0 \exp\left(-\frac{u}{D}x\right) \qquad (s1-2)$$

where $\rho_0$ is the particle density at $x = 0$. In the presence of a constant external force $f$, the potential energy of a particle at $x$ is $fx$. The statistical average of the potential energy per particle is described by

$$\frac{\int_0^\infty dx f x \rho_{st}(x)}{\int_0^\infty dx \rho_{st}(x)} = \frac{\int_0^\infty dx f x \rho_0 \exp\left(-\frac{u}{D}x\right)}{\int_0^\infty dx \rho_0 \exp\left(-\frac{u}{D}x\right)} = f\frac{D}{u}. \qquad (s1-3)$$

Within living cells, there exist both the random thermal force $f_T$ due to molecular collisions and the nonthermal force $f_A$ generated by ubiquitous molecular motors. The total force $f$ per particle is then

$$f = f_T + f_A. \qquad (s1-4)$$

Only the potential energy associated with $f_T$ follows the law of equipartition of energy, in which it equals the energy received from the surroundings as seen in Brownian motion. From Eqs. (s1-3) and (s1-4),

$$(f - f_A)\frac{D}{u} = k_B T \qquad (s1-5)$$

where $k_B$ is the Boltzmann constant, and $T$ is the absolute temperature.

FCS is a technique used to determine diffusion coefficients by analyzing fluctuations in the fluorescence intensity of molecules. The fluctuations observed in FCS arise from the combined effect of both thermal and nonthermal forces. Using parameters obtained with FCS, specifically the effective volume length $l$ and the correlation time $\tau_B$, which corresponds to the characteristic transit time of a particle through this volume, the average velocity $u$ is expressed as

$$u = \frac{l}{\tau_B}. \qquad (\mathrm{s1}-6)$$

From the macroscopic relationship between average particle velocity and force,

$$u = \mu f \qquad (\mathrm{s1}-7)$$

where $\mu$ is the particle mobility, indicating how easily particles move. From Eqs. (s1-6) and (s1-7), Eq. (s1-5) is rewritten as

$$f_A = \frac{l}{D\tau_B}\left(\frac{D}{\mu} - k_B T\right). \qquad (\mathrm{s1}-8).$$

The second term on the right side of Eq. (s1-8) represents the thermal fluctuation component. By measuring the deviation of observed fluctuations from those predicted for purely thermal fluctuations, we can determine the nonthermal forces. Note that in the first term on the right side of Eq. (s1-8), for a spherical particle, $\mu$ is determined from the Stokes-Einstein relation, and the other parameters are obtained from FCS data. In other words, analysis of the microscopic motion of particles subject to thermal and nonthermal forces provides an effective measure of active nonequilibrium driving in biological systems. Unlike previous studies, this approach does not require measuring passive responses of individual particles to externally applied mechanical stimuli.

## Derivation of the stochastic model for intracellular nonequilibrium fluctuations

The stochastic dynamics of a probe particle in the intracellular environment is described by a generalized Langevin equation. Within the viscoelastic cytoplasm, the particle experiences an elastic constraint $k$ and a viscous drag coefficient $\gamma$ within the viscoelastic cytoplasm. The equation of motion is given by:

$$\gamma \frac{dx(t)}{dt} + kx(t) = F_{th}(t) + F_{act}(t) \quad (s2-1)$$

where $F_{th}(t)$ represents thermal fluctuations and $F_{act}(t)$ denotes active forces driven by nonequilibrium processes.

Taking the Fourier transform of Eq. (s2-1), the displacement in the frequency domain is expressed as $\tilde{x}(\omega) = \chi(\omega)[\widetilde{F_{th}}(\omega) + \widetilde{F_{act}}(\omega)]$, where $\chi(\omega) = 1/(k + i\omega\gamma)$ is the complex response function. The displacement power spectral density $C(\omega) = \langle |\tilde{x}(\omega)|^2 \rangle$ is obtained by assuming that thermal and active forces are statistically independent. Substituting the response function yields the expression corresponding to Eq. (2):

$$C(\omega) = \frac{S_{th} + S_{act}(\omega)}{k^2 + \gamma^2\omega^2} \quad (s2-2)$$

According to the fluctuation-dissipation theorem, the thermal noise spectrum is a constant given by $S_{th} = 2\gamma k_B T$.

The active noise spectrum $S_{act}(\omega)$ is modeled as a superposition of multiple active processes with distinct characteristic time scales $\tau_i$, representing fast molecular motor steps and slow network remodeling. Assuming that each active component exhibits exponential correlations, the Wiener-Khinchin theorem yields the Lorentzian form used in Eq. (3):

$$S_{act}(\omega) = \sum_{i \in \{\text{fast,slow}\}} \frac{2\tau_i f_i^2}{1 + (\omega\tau_i)^2} \quad (s2-3)$$

where $f_i$ represents the magnitude of the active force at each time scale.

## Fluorescence correlation spectroscopy analysis

To probe intracellular dynamics, we utilized FCS, a technique that analyzes spontaneous fluctuations in fluorescence intensity, $I(t)$, from diffusing molecules within a microscopic observation volume. The statistical properties of these fluctuations were quantified using the fluorescence intensity time-autocorrelation function (ACF), $G(\tau)$, defined as:

$$G(\tau) = \frac{\langle \delta I(t)\delta I(t+\tau) \rangle}{\langle I(t) \rangle^2} \quad (s3-1)$$

where $\delta I(t)$ represents the fluctuation from the mean intensity, and $\tau$ is the lag time.

FCS measurements were performed on a confocal laser scanning microscope (Olympus FV3000) equipped with a 60× oil-immersion objective (NA 1.42). Live cells cultured on glass-bottom dishes for 24 h after transfection were maintained at 37 °C and 5% $CO_2$ using a stage-top incubator. The pinhole was set to 200 μm, defining a three-dimensional Gaussian observation volume with a radial dimension of $\omega_{xy}$ = 0.211 μm and a structural parameter of $\omega_z/\omega_{xy}$ = 2.67.

The autocorrelation function can be expressed in terms of the diffusion coefficient $D$ and the average particle number $\overline{N}$ as follows:

$$G(\tau) = \frac{1}{\overline{N}}\left(1+\frac{4D\tau}{{\omega_{xy}}^2}\right)^{-1}\left(1+\frac{4D\tau}{{\omega_z}^2}\right)^{-1/2}. \qquad (s3-2)$$

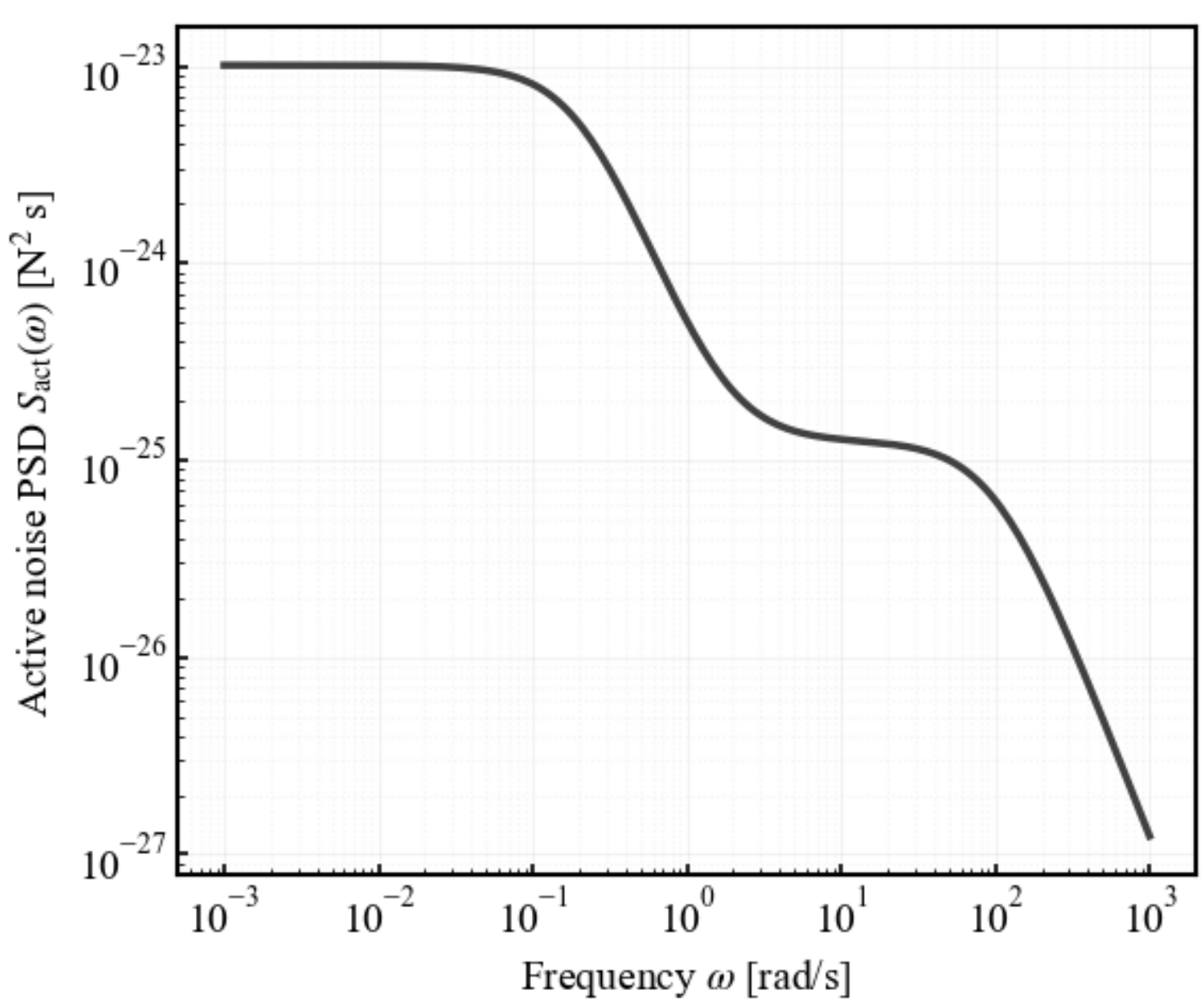

**Fig. S1 Theoretical profile of the multiscale active noise spectrum.**

The plot illustrates the active noise power spectral density $S_{act}(\omega)$, formulated in Eq. (3), as a function of frequency $\omega$. The spectrum exhibits two distinct plateaus corresponding to different characteristic time scales.

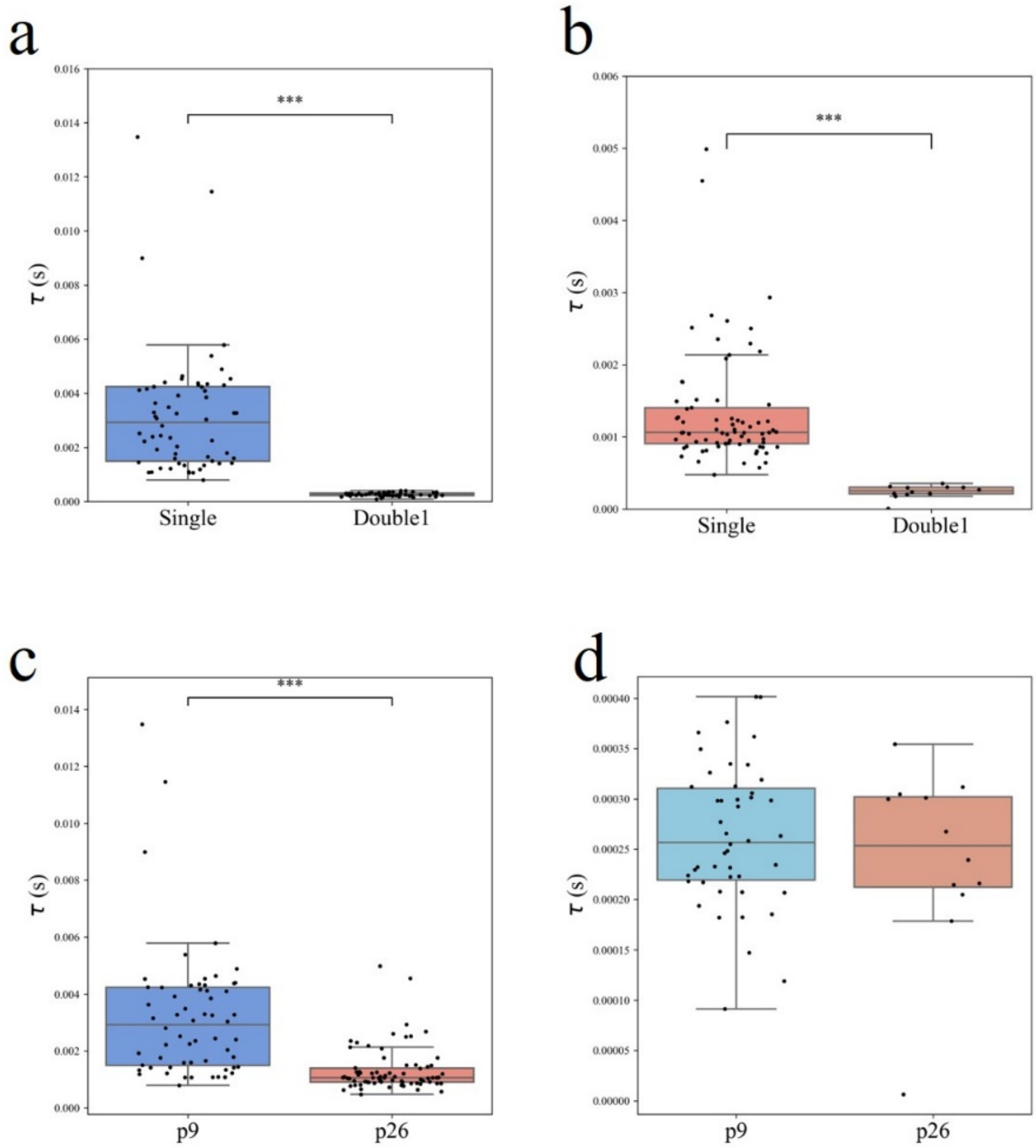

**Fig. S2 Cellular senescence alters diffusion heterogeneity.**

(a, b) Comparison of characteristic diffusion times between single- and double-fitting models shows significant differences, indicating that the multicomponent mode reflects a distinct mechanical complexity in the diffusion process. (c) Characteristic diffusion times for the single-mode category in p9 and p26 cells. (d) Characteristic diffusion times for the double-mode category in young (p9) and senescent (p26) cells.

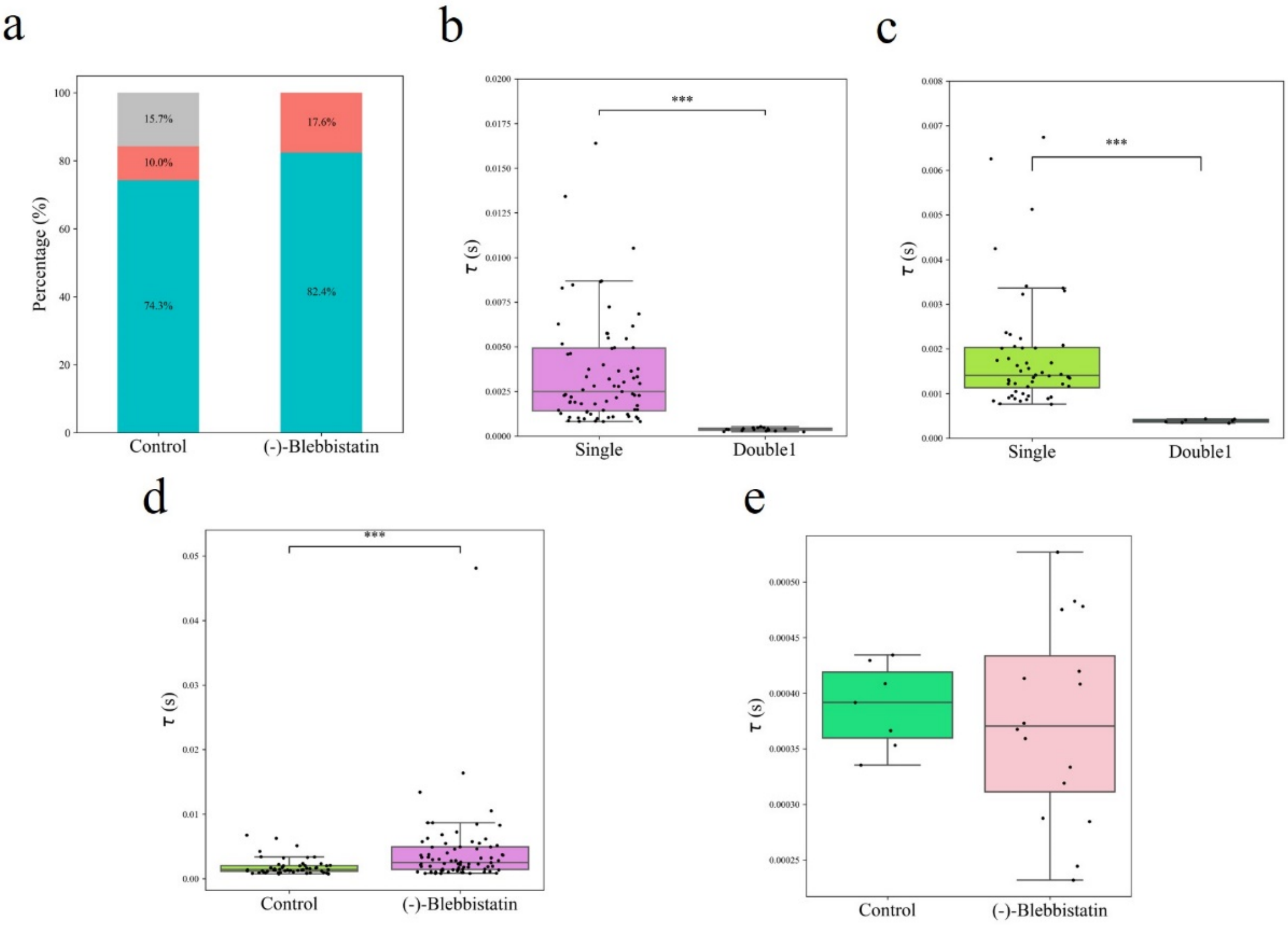

**Fig. S3 Myosin II activity regulates the likelihood of distinct diffusion modes.**

(a) Proportions of fitting modes in control and (-)-blebbistatin-treated cells. (b, c) Characteristic diffusion times for single- and double-fitting models under control (b) and (-)-blebbistatin-treated (c) conditions. (d) Comparison of characteristic diffusion times for the single-mode category between control and (-)-blebbistatin-treated cells. (e) Comparison of characteristic diffusion times for the double-mode category between control and (-)-blebbistatin-treated cells.